Abstract

There has been important understanding of the process by which a hypersonic dust impact makes an electrical signal on a spacecraft sensor, leading to a fuller understanding of the physics. Zaslavsky (2015) showed that the most important signal comes from the charging of the spacecraft, less from charging of an antenna. The present work is an extension of the work of Zaslavsky. An analytical treatment of the physics of a hypersonic dust impact and the mechanism for generating an electrical signal in a sensor, an antenna, is presented. The treatment is compared with observations from STEREO and Parker Solar Probe. A full treatment of this process by simulations seems beyond present computer capabilities, but some parts of the treatment can must depend on simulations but other features can be better understood through analytical treatment.

Evidence for a somewhat larger contribution from the antenna part of the signal than in previous publications is presented. Importance of electrostatic forces in forming the exiting plasma cloud is emphasized. Electrostatic forces lead to a rapid expansion of the escaping cloud, so that it expands more rapidly than escapes, and frequently surrounds one or more antennas. This accounts for the ability of dipole antennas to detect dust impacts.

Some progress toward an understanding occasional negative charging of an antenna is presented, together with direct evidence of such charging.

Use of laboratory measurements of charge to estimate size of spacecraft impacts are shown to be not reliable without further calibration work.

# Toward a Physics Based Model of Hypervelocity Dust Impacts

**Paul J Kellogg[1], S.D.Bale[2,3], Keith Goetz[1], Steven J. Monson[1]**

1 School of Physics and Astronomy, University of Minnesota, Minneapolis, 55455, USA
2 Space Sciences Laboratory, University of California, Berkeley, CA 94720-7450, USA
3 Physics Department, University of California, Berkeley, CA 94720-7300, USA

## 1. Introduction

Important progress has recently been made in understanding the process by which a hypervelocity dust particle impact makes an electrical signal on a spacecraft antenna (Aubier et al 1983, Gurnett et al 1983, Meyer-Vernet et al.2009, Meyer-Vernet et al 2014, 2017, Zaslavsky (2015), Veverka et al 2017). Understanding of this process now allows comparison of observations with understanding of the underlying physics. Important work in this direction has been done by Zaslavsky (2015). This work builds on Zaslavsky's, taken as a starting point. Some differences have been found, but the basic ideas of the authors quoted above represent a major advance. A similar extension of Zaslavsky's work has been published by Mann et al (2019).

The basic picture is that such a dust impact triggers the release of a cloud of ionized material which is initially neutral. Electrons are quickly removed from the cloud by two processes, both of which continue until the potential of the cloud prevents further escape. First, electrons have a larger thermal speed than the ions and so escape early.  Second, spacecraft in the solar wind at 1 AU are charged positively because photoemission of electrons exceeds the collection of ambient electrons by a considerable factor. The spacecraft is then charged positively until the resulting electric field returns enough electrons to balance the fluxes.  This electric field strips electrons from the cloud. At 1 A.U. the two effects result in the effective emission of a positive cloud, leaving the spacecraft negative.  The positive cloud may or may not interact with the antennas themselves, but in general the antennas tend to be maintained closer to the potential determined by their photoemission-pickup balance, and so do not change potential as much as the spacecraft does, resulting in a positive antenna potential relative to the spacecraft, as is observed.  The first escaping electrons sometimes make a short initial negative pulse.

An important addition to this picture was made by Zaslavsky (2015).  He found that the observed waveforms could only be understood if an appreciable number of electrons are deposited on an antenna, leaving the antenna negative, and providing a long tail of recovery to the observed waveform.  Much earlier work had considered deposition of charge from the emitted cloud, but had found the amount of charge deposited too small to account for the signal. In this work a source of this negative charge is suggested.

In this work, data from the monopole antennas (Bale et al. (2008))  on the STEREO spacecraft will be analyzed.  The two STEREO spacecraft are 3 axis stabilized and the three 6 meter long, mutually orthogonal monopole antennas are mounted on the side opposite to the sun.  A model of the waveform of a dust impact will be fitted to observed waveforms, and the fits used to evaluate various parameters of the impact, such as the potential rise time, the fraction of the signal due to antenna charging vs spacecraft charging, the restore times of the exponential returns to pre-impact conditions, etc.

## 2. Physics and Signal Waveform

### 2.1  Observations

In this work, the seminal work of Zaslavsky (2015) will be extended, with special attention to the shape of the dust impact waveforms, to try to derive further characteristics of the impacts.  The waveform of a dust impact which shows the characteristics to be analyzed, as seen on the three monopole antennas of the STEREO A spacecraft, is shown as Figure 1.  The signal is asymmetrical, with a rapid rise and a slower decay.  Attention is drawn to the signal on the X antenna, which decays past zero and then recovers.  This overshoot is not a universal feature of the impacts, but is reasonably common.  An explanation of this was given by Zaslavsky, who realized that some negative charge must have been given to the antenna.  The full explanation of how an antenna collects a negative change is not yet completely clear, but some suggestions will appear in Section 2.5.5.

The black line in Figure 1 is the result of a fit of a function which will be used to obtain many properties of the impact which are not measured directly.  The function to be fitted is discussed in Section 2.2.  Not all of the fits are this good, of course.  A normalized chi is shown as F% N_chi and deviations from the fit with N_chi larger than .2 have been rejected from the analysis.

The basic picture presented in the introduction implies that a major part of the signal comes not from a cloud striking the antenna but from the change of potential of the spacecraft. The signals then should be roughly the same on all three antennas, with some small differences due to the different couplings of an antenna to the ambient plasma, mainly due to different amount of shadowing by the spacecraft. However, if the impact is close to an antenna, the cloud will surround that antenna and cause a substantially different signal.

In Figure 2 is presented an initial confirmation of this picture. As the antennas are all on the anti-sunward side of the antenna, it should be expected that at least half of the signals result from impacts on the sunward side which are distant from any antenna. In Figure 2 are shown the ratios of the peak signals of pairs of antennas. It will be seen that a significant number of these ratios are concentrated near unity but there are also some large differences from unity, in accordance with the picture presented above. As mentioned, the large differences are most probably due to immersion of one or more of the antennas in the cloud released by the impact.

To begin a more detailed discussion of a dust particle impact and subsequent signal on an electric antenna, it is necessary to start with the physics of the impacting dust particle and the target material. A complete discussion is too complex for this elementary treatment, and would require extensive simulation, but some basic ideas can be considered. Two simulations will be extensively quoted here, Samela and Nordlund (2010), and Fletcher et al (2015). There are other simulations, (e.g. Gong et al 2019) some more advanced, but extensive simulations carried out to answer some of the basic physics questions raised here are not known to the authors.

The Samela and Nordlund (2010) simulation uses a particle in cell procedure, but the target is not charged so there is no stripping of electrons from the emitted cloud which is therefor also uncharged. In fact, the molecules are not ionized and there are no electrons. Fletcher at al (2015) use a sophisticated hydrocode, including charges, but apparently there is no charge on the target. Therefor neither of these simulations can properly represent the full physics of the impact according to the understanding of the physics presented here. However they should give some understanding of the early impact and effect on the target,

## 2.2. Model Fit to the Observations

The measured signal, on the present picture of the interaction, has two parts, the negative change in potential of the spacecraft, and a change of potential of the antenna, the observed signal being their difference. Observations like those of Figure 1 are fitted to a model. The model has 6 variables X[0] to X[5] and the observed shape is fitted by:

$$V[nf] = X[0]*(1.-exp(-t/X[2])) * (exp(-(t/X[3])) -x[5]*exp(-t/x=X[4])) \qquad (1)$$

where V is the antenna potential relative to the spacecraft, nf is the observed sample number, X[1] is the start time of the signal in ms after the start of the event sampling, t is the time in ms after [X[1], X[0] is a signal amplitude in mV, and X[1] is the variable $Q/(4 \pi \varepsilon_0 R)$, X[2] is its rise time in mv/ms. X[3] and x[4] are the exponential restore times, X[5] is the fraction of the signal from the antenna. An example of such a fit to a dust impact on 08 Jan 2010 0357:41.992 is shown as Figure 1, which shows the signal as seen on each of the three antennas. The measured signals are shown in color, and the black lines show the results of the fit.

The overshoot in the antennas has been difficult to understand. Zaslavsky (2015) pointed out that cases of extreme overshoot imply that some negative charge has been added to an antenna. This charge probably comes from the stripping and electron escape period, when the impact cloud has not expanded very much and the electrons are dense. The importance of this effect was not understood when the study of STEREO data reported here was done, and the model does not include effect of negative charge delivered to the antennas.

## 2.3 The Impact and Crater Formation

To begin, the dust particle hits the target and penetrates. As the impact speed, v, considered here is well above the speed of sound in either the dust particle or the target, the stopping is only due to the acceleration of the target material to the speed of the dust particle. This implies a stopping pressure of $\rho_T v^2$ where $\rho_T$ is the density if the target. This very high pressure at the front of the penetrating projectile also implies a sideward force. This is shown by a cavity made in low density material, where the reductions of the sideward pressure is demonstrated by the observed carrot shaped cavity, narrowing as the projectile slows (Burchell et al 2008, Westphal et el 2014). Two conclusions follow: first, the penetration depth depends on the ratio of the target density to the projectile density, and second, the sideways force delivers much energy to the target.

This sideward pressure then starts a shock wave into the target so that a large volume is affected by the impact. Measurements give a ratio of cavity volume to projectile volume showing that for nearly equal densities the cavity is not a cylinder, but is nearly a hemisphere. This demonstrates the importance of compression force to the sides.

The volume of the crater produced by dust impacts in space is large compared to the volume of the projectile (e.g. Shanbing et al 1994). Burchell et al (1999) have made a careful measurement, and find, as do some others, that the volume is roughly proportional to the kinetic energy of the projectile. Their fit to their measurements give, for the crater volume Vc and impact speed v

$$Vc = 0{:}00256 \, v^{2.11}$$

Here Vc is the crater volume in cm$^3$ and v is the impact speed in km/s. These formulas imply that at an impact speed of 60 km/s the ratio of crater volume to the impactor volume would be 28,000. Note that the Burchell et al fit gives a power of speed slightly greater than 2, as did Mocker et al (2013) for high speeds.

The coupling of a dust impact to an electric antenna must, obviously, involve the charge created by the impact. A long series of laboratory measurements have shown that the charge created is proportional to velocity to a higher power than just the energy, a power of v 3.5 to 4.5 (Smith and Adams (1973), Iglseder and Igenbergs (1990), Goldsworthy et al, 200**3**), Mocker et al (2013) and references therein.. It is also known that incoming interstellar dust, whose velocity relative to a spacecraft is increased by the gravity of the sun when it is opposed to the orbital velocity of the STEREO spacecraft, up to 79 km/s, causes an increase in dust impacts which is statistically recognizable. Zaslavaky et al (2012), Kellogg et al (2016). That is why we assume that most of the observed impacts during the rest of the year, which appear to be similar, are due to retrograde dust, with a velocity of 60 km/s

If all of this matter were ejected with the cloud and if the cloud had time to reach equilibrium, it would not be fully ionized. However, an ionized cloud is observed in the laboratory and seems necessary to make an electric signal on an antenna in space. Simulations agree and show that only a small amount of target material, perhaps less than the volume of the projectile, is ejected. Further, a shock wave is generated in the target, which apparently causes ejection of material, which must be target material, somewhat later than the primary ejection of the projectile material. A possible picture of the impact process is then that the highly compressed impact particle is expelled immediately, with a small amount of target material, and then later the shocked target material is expelled. This latter material must be only weakly ionized if at all, and is perhaps not even atomized but is expelled as chunks. This may be what is observed (St.Syr et al, 2009, Thompson et al 2009).

The charge released is proportional to a power of the projectile speed higher then 2, usually between 3 and 4.5. This presents a puzzle. The crater volume is proportional to the particle kinetic energy. Further, the simulation of Fletcher et al (2015) finds that the outgoing cloud is 100% ionized. On the other hand, the experiments find that the production of charge is proportional to a higher power of velocity than 2, i.e. between 3 and 4.5. If the cloud is 100% ionized at low velocities, where does the increase at higher velocities come from? It follows that either not all of the material is ionized when the impact speed is in the laboratory range or that some ions must be multiply ionized at speeds relevant to spacecraft impacts. Multiple ionization seems probable.

## 2.4 Expansion from the Cavity

At the bottom of the cavity there is then a mass of combined projectile and compressed target material. We skip the details of emergence of the material from the cavity to start with a mass composed of the impacting particle plus some target material that previously occupied the cavity, with this material just outside the cavity, expanding, and with a center travelling away from the target at the thermal velocity. Simulations are required to determine the division between target and projectile material and the energy which has been delivered to the target. This is beyond the simple physics to be used here, and so some guesses will be made to allow further progress. It seems that the compressed material originally in the target should form part of the expelled mass but this is not in accord with the simulation of Samela and Nordlund (2010) who show that only a small amount of target material is expelled. The tables below present some guesses as to this division. A particle arriving at 60 km/s, the approximate speed of retrograde dust and close to that of interstellar dust, arrives with an energy of 18 eV/nucleon. Here three guesses are that the particle has lost half of its energy, 90% and 97% of its energy to the target. These three guesses are based on the presentations of the simulations above and on the observations of Ratcliff and Allahdadi (1996), who found that only 3% of the impactor energy was used to ionize the cloud. Further, 20% of the projectile mass from the target has been added to the material which is going to exit rapidly. It is also assumed that recombination can be neglected during the exit. The material will be ionized and multiply ionized in the 50% case. Hence a somewhat expanded clump of material arrives just beyond the exit of the cavity. Again guessing that the volume has expanded adiabatically by a factor of 10, the density of this clump will be about .1 gm/cm$^3$. Some results of these guesses are shown in Table I below. This table is estimates for initial (before impact) spherical particles of 10, 1 and 0.1micron diameter. The atomic weight has been taken as 18 and the thermal velocity is that of singly ionized ions.

Table I

| Pre-impact Diameter($\mu$) | post-impact diameter($\mu$) | T (eV) | Vth (km/s) | $\phi$ (V) | E (V/m) | final dia (m) |
|---|---|---|---|---|---|---|
| 50 % Energy to target | | | | | | |
| 10 | 12.6 | 134. | 38. | $5\ 10^8$ | $1\ 10^{14}$ | 240. |
| 1 | 1.26 | 134. | 38. | $5\ 10^6$ | $1\ 10^{13}$ | 24. |
| 0.1 | .126 | 134. | 38. | $5\ 10^4$ | $1\ 10^{12}$ | 2.4 |
| 90 % Energy to target | | | | | | |
| 10 | 12.6 | 27. | 17. | $5\ 10^8$ | $1\ 10^{14}$ | 240. |
| 1 | 1.26 | 27. | 17. | $5\ 10^6$ | $1\ 10^{13}$ | 24. |
| 0.1 | .126 | 27. | 17. | $5\ 10^4$ | $1\ 10^{12}$ | 2.4 |
| 97 % Energy to target (Ratcliff and Allahdadi (1996)) | | | | | | |
| 10 | 12.6 | 8. | 6.5. | $5\ 10^8$ | $1\ 10^{14}$ | 240. |
| 1 | 1.26 | 8. | 6.5 | $5\ 10^6$ | $1\ 10^{13}$ | 24. |
| 0.1 | .126 | 8. | 6.5 | $5\ 10^4$ | $1\ 10^{12}$ | 2.4 |

The last two columns are the surface potential and surface electric field on the assumption that the stripping is complete, i.e. all ionized electrons have been stripped and the cloud consists only of the ions. It is clear from these very large estimates that the stripping cannot be complete, and that only a small fraction of the electrons can be removed from the cloud.
  In the laboratory, the stripping field, (which is however often of the sign to strip ions) is generally of the order of kilovolts/m, while in the solar wind it is only of the order of 10 V/m. It follows that the laboratory measurements of emitted charge as a function of impact speed cannot be used directly, but would require some additional calibration procedure. The dependence of charge on the stripping field has been demonstrated and investigated by Lee et al (2013).

**2.5 Escape from the Spacecraft**
  Then there is a highly compressed mixture of the impacting particle and the target material. This mass then expands and moves away. The expansion is partly thermal and partly driven by the electrostatic self-repulsion of the ions. Both the expansion speed and the escape speed will initially be of the order of the thermal speed, shown in the table above. The two simulations used here differ strongly on these speeds. Samela and Nordlund (2010), considered the impact of a cluster of 114534 Argon atoms with silicon at a speed of 22 km/s, therefor at kinetic energy 100 eV/atom. The expansion speed, scaled from their Figure 3 (2010) is about 1 km/s for the outer edge of the cloud and .75 km/s for a dense part, obtained from measuring their Figure 3 at 20 ps after the impact of the Argon cluster. The Fletcher et al (2015) simulation shows a compilations of impact speeds for iron particles on tungsten and detail for an impact at 20 km/s. For the compilation, the ratio of expansion speed to impact speed is about 0.7 for a (small) 1 ng particle, up to impact speeds of 50 km/s. It is to be noted that the Fletcher et al (2015) simulation finds that the temperature of the cloud is a constant 2.5 eV independent of impact speed for large speed, while laboratory measurements (Miyachi et al (2008), Collette et al (2016)) find that the temperature is proportional to the square root of the impact speed.

As the simulations referenced above seem to have neglected charging it then expands adiabatically at the thermal rate. However, it seems that in practice, the electrostatic self-repulsion plays an important role, so both expansion processes will be considered. For adiabatic expansion, the expansion rate decreases with the temperature but the escape rate does not so that in the adiabatic case the expanding cloud is small compared to the distance from the spacecraft, as is shown by the simulation of Fletcher et al (2015) and by the estimates below.

The rise time of the signal (Figure 1) can be interpreted to give information on the expansion and ejection speeds. The data, obtained from the fits mentioned in the discussion of Figure 1give expansion speed, or rather signal rise speed, in mV/ms. Conversion to length/sec is to be discussed in Section 2.5.2

**2.5.1 Adiabatic expansion**

The two simulations referenced above, Samela and Nordlund (2010), and Fletcher et al (2015), are expansions from an uncharged cloud. Therefor the expansion is adiabatic, driven by the thermal energy of the cloud. It is of interest to consider adiabatic expansion, though it will be shown that the expansion will be more rapid due to the mutual repulsion of the charged particles.

Adiabatic expansion implies that the radius of the cloud will increase at a rate $\sqrt{k_B T/M}$ where M is the effective mass of the ions, which has been argued above is mainly the mass of the atoms of the projectile. The temperature follows the adiabatic monoatomic law

$$dR/dt = \sqrt{k_B T/M}, \text{ implying } T \alpha 1/R^2$$

Giving $dR/dt = \sqrt{kT_0/m}\, R_0/R$ and $R^2 = R_0^2 + 2\sqrt{k T_0/m}\, t^2$

The same relation can be obtained by considering the energy. The work done by the pressure in the expansion is pdV In the work following it will be found that the expansion is partly driven by the self repulsion of the positive charge. The separation of pressure work and electric field work could be used to determine the cooling in this more complicated case but this has not been done here.

It was thought to continue the adiabatic expansion of the cloud until the mean free path (L) of the particles is of the order of the size of the cloud and the cloud becomes collisionless. A curious result of adiabatic expansion is that the cross section increases so rapidly that this does not happen and the cloud remains collisional. Assuming that Coulomb collisions dominate the process, the time for equilibrium, the "self collision time" of Spitzer (1956) is

$$t_c = M^{1/2} (3k_B T)^{3/2}/(5.712 \pi n e^4 Z^4 \ln \Lambda)$$

(n is the number density) and using the usual $L \sigma / v$ = collision time with $L = v t_c$ and $v = (k_B T/M)^{1/2}$ gives a mean free path, L, of:

$$L = (k_B T/M)^{1/2} t_c = (3)^{3/2} (k_B T)^2/(5.712 \pi n e^4 Z^4 \ln \Lambda)$$

Replacing $T = T_0 R_0^2/R^2$ and $n = n_0 R_0^3/R^3$ gives L proportional to $R^{-1}$. It can be seen that the mean free path decreases as R increases. At the start, obviously, the mean free path is much smaller than the size of the nearly solid cloud, and it continues to decrease as the radius increases, so they are never equal. Therefor some other consideration must determine the collisional-noncollisional transition.

This odd conclusion comes from the fact that the collision cross section increases so fast as the thermal velocity decreases. It does not make sense to have a cross section which is larger than the spacing between scatterers, so replacing $\sigma$ with $(.5)^2/n^{2/3}$, derived from the distance between scatterers gives:

$$L = 4/n^{1/3}$$

As R has the same dependence on n, this does not give a reasonable answer either.

### 2.5.2. Observed expansion and escape, method of images

In view of this large difference in escape speeds of the two simulations, some data from impacts on the STEREO spacecraft are presented here, in Figure 3. The impact speed and the nature of the impacting dust are unknown, and the details of the interaction of the charged cloud with the antenna are uncertain, but the difference above, a factor of 14, is probably large compared to these uncertainties.

The rise time of the signal (Figure 1) can be interpreted to give information on the expansion and ejection speeds. The data, obtained from the fits mentioned in the discussion of Figure 1 and described in Section 2.2, give expansion speed, or rather signal rise speed in mV/ms. To convert to length/sec, the following picture has been used. When two oppositely charged bodies are close together, the potential of each is reduced. Thus the potential rise rate is taken from the problem of a point charge near a sphere, solved by the method of images (Jackson 1999). The point charge is presumed to be at the center of the ionized cloud here, but the result holds for a spherical cloud also. The potential of the sphere of radius R when the point charge Q is at a distance H from the surface of the sphere is:

$$V = -Q \, (H/(H+R))/(4 \pi \varepsilon_0 R) \qquad (2)$$

Although the spacecraft is not a sphere, it seems reasonable to take R to be half the mean value of its three sides, .80 m. To evaluate Q, the rising part of the signal is fitted to the function described in Section 2.2. The rising speed, dH/dt is then related to the observed rising speed in V/s through:

$$dH/dt \ (m/s) = ((4 \pi \varepsilon_0 R)/Q) * (H+R)^2/R) \ dV/dt \quad (mv/ms) \qquad (3)$$

When H is small compared to R the relation is:

$$dH/dt \ (m/s) = ((4 \pi \varepsilon_0 R)/Q) * R \ dV/dt \quad (mv/ms) \qquad (3a)$$

and when the cloud is very far away the potential would be:

$$V = Q/(4 \pi \varepsilon_0 R) \qquad (3c)$$

The parameter $Q/(4 \pi \varepsilon_0 R)$ is evaluated as part of the fit described in Section 2.2.

In Fig 3 is shown a histogram of expansion and ejection speeds from this model. The upper panel shows measured rates of rise of the signal, in mV/ms. These rates are obtained by fitting the model of Section 2.2 to measurements of antenna potential as a function of time. In the lower panel of the figure the estimated rise speed in km/s is shown. In this panel the limit H<<R, is used to convert the measured speed in mV/ms to km/s, as above. Note, however, that if the full dependence shown in Eq (3) were to be used, then a uniform rise with dH/dt constant would appear to be slightly slowed in dV/dt as the rise progresses. This will be important when acceleration in the spacecraft field is discussed in Section 2.3.2, and tends to reduce the observed acceleration.

This interpretation and the histogram of Figure 3 show an expansion speed between the values of the two simulations quoted. However, it must be noted that the expansion is by no means uniform in speed. If the expansion were to be driven adiabatically by the particle temperature, the temperature would decrease rapidly. As will be shown, it is more probably driven by the electric field of the positively charged cloud, but there will still be some cooling due to work done by the thermal energy. It is not clear that these data are completely immune from the fast expansion due to electrostatic forces, to be treated in Section 2.5.4. However there are few expansion speeds less that 10 km/s, and a large concentration near that speed. From Table I, it will be seen that 10 km/sec coincides with an energy transfer to the target of between 3 and 10 percent of the projectile kinetic energy.

It is usually considered that expansion continues until the density of the cloud equals the density of the ambient medium, in this case, the solar wind at a density of about 5 /cm$^3$. For an initial dust particle, taken as a sphere of 1 micron diameter and a density of 1 gm/cm$^3$, average atomic weight A of 18, this means expanding to a radius of about 5 meters. If the expansion were adiabatic, this means that the temperature would decrease by a factor of about $10^6$. This seems unreasonable and contrary to what is observed in the laboratory experiments (Miyachi et al (2008), Collette et al (2016)) and so the expansion must be driven by something other than thermal.

It will be noted that, in this adiabatic case, the escape speed remains constant while the expansion speed decreases as the cloud cools. This is in accord with the simulation of Fletcher et al (2015) which shows a cloud which remains small as it leaves the target. However, the Fletcher et al simulation is for a cloud escaping from an uncharged target, and it will be shown that the picture is changed for a charged target and cloud.

### 2.5.3 Stripping and Charging

The cloud next loses some electrons, some to the spacecraft, for two reasons. First, the electrons of the cloud are faster than the ions and so some will leave the cloud. Second, as the cloud crosses the electric field of the spacecraft, some of the electrons will be attracted back to the spacecraft by this field. Both processes will deliver electrons to the spacecraft and to the antennas. As shown in connection with Table I, this stripping cannot be complete.

Figure 1 shows the waveform for a typical impact at 1 AU, where the spacecraft is charged. There is a slight negative dip in the fitted curve but this is an artifact due to a slight error in the start of the event. However similar dips, though rare, are sometimes seen on STEREO (Kellogg et al, 2016, Fig. 4). During the first perihelion of Parker Solar Probe, the spacecraft potential was nearly zero, as had been predicted by Ergun et al (2010) as due to secondary emission caused by the higher ambient electron temperature. In Figure 4 is shown a Parker Solar Probe waveform measured by the Time Domain Sampler (TDS) of the Fields experiment (Bale et al. 2016). This event shows what we, and others (Collette et al 2015, Mann et al 2019) interpret as an initial electron burst due to the higher thermal speed of electrons, the negative signal between 4.23 and 4.25 ms. Unlike STEREO, for Parker Solar Probe close to the sun these negative pulses are nearly universal. The waveforms transmitted by the TDS during this first perihelion are wildly variable, and this example has been picked for its resemblance to waveforms at 1 AU. At this first perihelion at about 35 solar radii, the speed of a particle in a circular orbit is about 180 km/sec, the speed of the spacecraft is larger, and so, given the sensitivity of the emitted charge to impactor speed, variations from the 1 AU data are to be expected. The difference between the rarity of such STEREO events and their near universality on PSP is perhaps due to the large difference in impact energy. Among other reasons, since multiple ionizations are likely even at 60 km/sec, a considerable contribution from highly ionized atoms might be expected, with more electrons.

It seems that the first negative spike must be due to the fast escape, and must be due to such electrons striking the antenna, as increasing negative charge on the spacecraft should produce a positive signal. Then what is the overshoot due to, as it is presumed that it also involves addition of negative change to an antenna? It is proposed that the structure of electron emission is much more complex that the simple model used in Section 2.5.4, in that the initial negative spike is fast electrons escaping until the cloud is sufficiently charged that its potential forbids further escape, and the second, the overshoot, is due to further escape as the cloud expands and its potential decreases due to its increasing radius allowing further escape, perhaps from a different structure of the cloud.

### 2.5.4 Electrostatic driven expansion

To properly understand the behavior of the cloud after it has left the cavity it is necessary to take into account the mutual repulsion of the predominating ions. The positive charge in the cloud drives an expansion that increases the rate beyond adiabatic expansion. The first charging of the cloud is from the escape of fast electrons until the potential of the cloud is of the order of 10 or a few 10's of volts. For the recently formed cloud of radius microns, the electric field is of the order of megavolts/m, a field which should not be ignored. This electrostatic expansion is complicated and only a very rough approximation will be attempted here with the purpose of showing that electrostatic expansion is more rapid than adiabatic expansion, but a complete description must be left to later work. First, as the newborn charged cloud begins expansion, the electrons will be redistributed. Electrons will seek the area of least potential energy on the time scale of the conductivity, very rapidly, which implies that they will not be uniformly distributed through the cloud, but rather will concentrate toward its center. So the central part of the cloud will be nearly neutral, and the positive ions will be concentrated in a surface layer. This is shown or discussed in some simulations (Peano et al 2007, Pantellini et al, 2012). This layer will expand rapidly, leaving a core of more neutral plasma which plays a smaller role in making a signal. Further, the temperature of the cloud will decrease, as some of the expansion is driven by the thermal energy of the particles, reducing particle energy. Here, however a very simple calculation is presented with the aim of showing that the expansion is faster than adiabatic expansion and faster than the escape speed. It is assumed that the cloud expands as a sphere, and that as the radius increases and the potential would tend to decrease, electrons escape to maintain a constant potential V, though in fact it would decrease as the electrons cool. It is also assumed that the charge distribution maintains a surface electric field that would be that of a sphere, i.e $E = V/R$, where R is the radius,

dRdt[0] = v

; the expansion speed is initially thermal and expands by force of electric field

dvdt = e*E_field/(A*Mp)

the distribution of the outer layer   then is accelerated outward according to

DR/dr = v   and dv/dt = eE/(A Mp) = eV/(R A Mp)

The results of integrating these equations, using V = 20 Volts are shown in Figure 5. The upper panel shows some detail during the initial acceleration and the lower panel shows the expansion to a radius of about 1 meter. The expansion to 1 meter takes only about 10 μs. It will be seen that the expansion overwhelms the escape, so that the cloud remains close to the spacecraft, in contrast to the case for adiabatic expansion and the simulation of Fletcher et al (2015).   It is clear that most of the acceleration of the cloud occurs early.  In the computation shown, the speed at reasonable distance is about 40 km/s.  This is too fast to agree with the observations reported in Section 2.5.5.  Lee et al (2011) measured the expansion speed in laboratory experiments and found expansion speeds from 10 km/s and above, in agreement with Section 2.5.4, and in mild disagreement with this result.   Some work was done on more sophisticated expansion calculations, taking into account slowing of the part of the expansion due to cooling and that the cloud potential phi is determined by the electron thermal energy It was then realized that the important parameter in these calculations is the radius of the cloud when the fast electrons escape, as the acceleration in that case is from E = phi/R.

### 2.5.5  Overshoot and Occasional Negative Charging of the Antennas

Returning to STEREO data, Figure 1 was chosen especially to show a large overshoot, which does not always occur but is reasonably common.  Zaslavsky attributed large overshoots to deposit of electrons on the antenna, and it seems likely that this deposit is from the electrons expelled in the initial periods when electrons are escaping profusely.  Figure 9, data from STEREO, shows two unusual events which seem to support this.  The overshoot is early and especially large. In each case, on one antenna the initial rise is quickly terminated and the signal becomes negative.  The proposed interpretation is that, after an initial rise due mainly to charging of the spacecraft, more electrons escaping from the cloud are deposited on the antenna. This is interpreted as a larger contribution from the spacecraft, which decays more rapidly (Zaslavski 2015), accounting for the sharper peak and leaves the signal from the antenna, of opposite sign. The time sequence of the overshoot suggests that the deposition of electrons is due rather to deposition of electrons from a cloud which has surrounded the antenna, rather than due to the initial faster electrons.  The fit of Section 2.2 gives x[5], the ratio of the antenna signal to the spacecraft signal to be .24 for X and  .14 and .18 for X  and antennas Y and Z respectively. These are somewhat larger than obtained by Zaslavsky (2015). The amount of charge so deposited is apparently quite large and so the impact must have occurred close to the antenna bases. In the data examined from STEREO A, all events from 2010 Jan 3 to 29 inclusive,164 impact events which were not saturated were analyzed and an approximately equal number were saturated and not kept. Therefor 2 out of about 330 impacts produced this unusual waveform.  The effective area of STEREO is about 21 m$^2$ (Kellogg et al 2009) as so this suggests that the impacts were in an area of 2/330 of 21 m$^2$ or within 20 cm of the antenna. In these two events the antenna rise is overcome and starts turning negative at about .05 ms after the impact and the negative signal reaches a maximum of the order of 3 ms after the impact. The first, with the 20 cm estimate, suggests an expansion speed of the order of 4 km/s. This is slower than the electrostatic expansion speed discussed in Section 2.5.4, and suggests that the guessed size when the fast electrons escape should be three times larger.

After noting these two extreme events, it was recognized that milder events, events in which one antenna signal is sharper at the top and drops faster and more than the others and even past zero as in Figure 1, frequently occur. These effects were not recognized when the fitting program of Section 2.2 was created, and the fit function does not properly take negative charging of an antenna into account.

The short negative pulse seen at the beginning of the Parker Solar Probe impact, Figure 4, is probably also due to the escape of electrons because they are faster.

## 2.6 Acceleration

As the positively charged cloud leaves the surface of the spacecraft it will be accelerated outward by the field of the spacecraft. This acceleration is not very large compared to the initial speed provided by the cloud temperature, but is sometimes visible. Further, Eqs 3 and 3a show that even a uniform rise with dH/dt constant would show curvature showing an apparent slowing of the rise in the observation of dV/dt. An example of the rise showing a curvature toward faster rise and outward acceleration is shown in Figure 6. This provides evidence for the general picture of dust impact signals as being due to emission of a positive cloud, as described in the Introduction.

## 2.9. Recovery

### 2.9.1 Calculated rates -- Expected Restore Times

The spacecraft and the antennas are restored to their equilibrium potentials through photoemission of electrons and capture of ambient ions and electrons. For collection of ambient electrons the rates calculated by Mott-Smith and Langmuir (1926) are used. For photoemission the rates of Pedersen (1995) are used. The recovery rates are

$$dV/dt = (I_{ph} + I_{amb})/C = (1/C)\, dI/dV\, \Delta V = 1/(R\, C)\, \Delta V$$

where $I_{ph}$ and $I_{amb}$ are the respective currents and R is the object-plasma resistance. The restore rate for antennas requires the Mott-Smith and Langmuir (1926) equation for cylinders, their Eq. 28. It will be seen that the restore time is just the usual RC time, the C being the capacitance of the spacecraft, taken to be the capacitance of an ellipsoid, 160 pF, (Kellogg et al 2018a,b) and the resistance being the antenna resistance, taken to be the derivative of the Mott-Smith and Langmuir current with respect to voltage. For the spacecraft, the corresponding current equation for a sphere is used, their Eq 36. Figure 7 shows the calculated resistance as a function of ambient density according to these current calculations, and Figure 8 shows calculations of the restore times. In these, the illuminated and total areas have been taken from Kellogg et al (2009) except that the antenna illuminated area has been multiplied by sin(125 deg) to take into account the slant of the antennas with respect to the sun direction. There is some question here, as to how to account for the high gain antenna and for the solar arrays. For the solar arrays, the sunward side, the photocell side, is insulated from the spacecraft and only the back, shadowed, side provides current from the electron thermal current. The currents from the solar arrays were discussed in Kellogg et al (2009), and an average of the two orientations presented there has been used as an estimate for the orientation during 2010 when the high gain (telemetry) antenna angle to the sun was intermediate.

The densities were obtained from STA_L2_PLA_1DMAX_1MIN on CDAWeb and from Antoinette Galvin. As the surfaces used by Pederson and others give difference functions for photoemission, the relative antenna potentials during periods of different densities were checked. The spacecraft-antenna potential differences calculated as above agreed very well for antennas X and Y, but not well for Z. Z has always been maverick. It is thought that this is because the shadowing of Z by the spacecraft results on a potential very close to that of the spacecraft and so other effects, especially electron temperature, are important. At any rate, no better photoemission parameters have been found.

### 2.9.2 Observation

In Figures 8 are also shown the measured restore time obtained from the fit described in Section 2.2, shown as diamonds. It will be seen that the measured restore times for the spacecraft are generally longer than the calculations, calculations based on established physics. The restore times for the antennas seem to bear no relation to the calculation. The calculations are based on the supposition that the concerned objects are in an ambient plasma with solar illumination. However the electrostatic expansion times found in Section 2.5.4 show that the impact cloud expands sufficiently rapidly that a major part of the expanding cloud remains near the spacecraft, and the positive plasma of the cloud still surrounds the elements of the system. It therefore seems that the difference between these observations and the calculations indicates an error in the assumed physical situation and indirectly supports the assumption that both systems are still immersed in the emitted material.

## 3. Summary and Conclusions

An attempt is made to treat the physics of how a hypervelocity dust impact makes an electric signal on an antenna in a spacecraft using analytical methods and results from observations, principally from STEREO and Parker Solar Probe. The basis of the approach used here is taken from Zaslavsky (2015). A full description of the process would require simulations, difficult simulations which may be beyond the capacity of presently available computers, but some guiding results are obtained. These include evidence that the largest part of the signal is due to the change of potential of the spacecraft, not of the antennas. Suggested are cautions on using laboratory emitted charge calibrations to determine the properties of the dust, importance of electric self-repulsion in the expansion process and discussion of surroundings of the spacecraft-antenna system after an impact.

The signal rise speed, used as a measure of the thermal energy in the cloud just after it has exited the impact cavity, indicates that only a small amount, between 3 and 10 % of the impactor kinetic energy, is taken by the escaping cloud and that the rest is transferred to the target. This target energy then feeds a shock wave which probably causes later emission of lower energy material.

It is found that the stripping of electrons from the cloud as it moves away from the target through an electric field cannot be complete and therefore will depend in the strength of this electric field, as was found by Lee et al (2013) in laboratory measurement. This must be taken into account if the laboratory measurements of charge are to be used.

It is found that, after the impact, the charged cloud expands rapidly under the self-repulsion of the ions, and therefore the cloud reaches a fairly large size before it has travelled far from the impact, thus frequently enveloping one or more antennas.  This demonstrates the validity of use of differences in antenna potentials to determine the position and therefore direction of the impact (Malaspina et al 2014) and also bears on the detection of dust impacts on spacecraft using dipole antennas  (Meyer-Vernet et al 2014,  Kellogg et al 2016)).

## Acknowledgements

The authors thank CDAweb and from Antoinette Galvin. (University of New Hampshire) for particle density and particle pressure data.  This work was supported by NASA grant NNX14AK73G, NASA contract NNN06AA01C and SEcurian Financial.
The data used, principally the data in Figures 1, 6, 9  and 10, are given in "supporting data".

## Figure Captions

**1.**  The waveform of a dust impact as measured on the three STEREO antennas.

2.  Ratios of the peak signals in pairs of antennas.

3.  Histogram of rise speeds.  See text for conversion from potential to rise in km/s.

4. Waveform of a dust impact on Parker Solar Probe near first perihelion at 35 Solar radii.

5.  Expansion speed due to electrostatic repulsion.  The upper panel shows the early acceleration    and the lower panel shows expansion to about 5 meters radius.

6.  Acceleration due to the spacecraft electric field acting on the positive charge of the cloud, an event on Ahead at 2010/01/17 2113.

7.  Calculations of the spacecraft-plasma and antenna-plasma resistance as a function of ambient plasma density.

8  Restore times as functions of density from the resistance calculations shown in Figure 7.

9. Two STEREO events thought to show deposition of electrons on an antenna

10. A similar event from the first perihelion of Parker Solar Probe.

Zaslavsky, A., (2015), Floating potential perturbations due to micrometeoroid impacts: Theory and application to S/WAVES dat

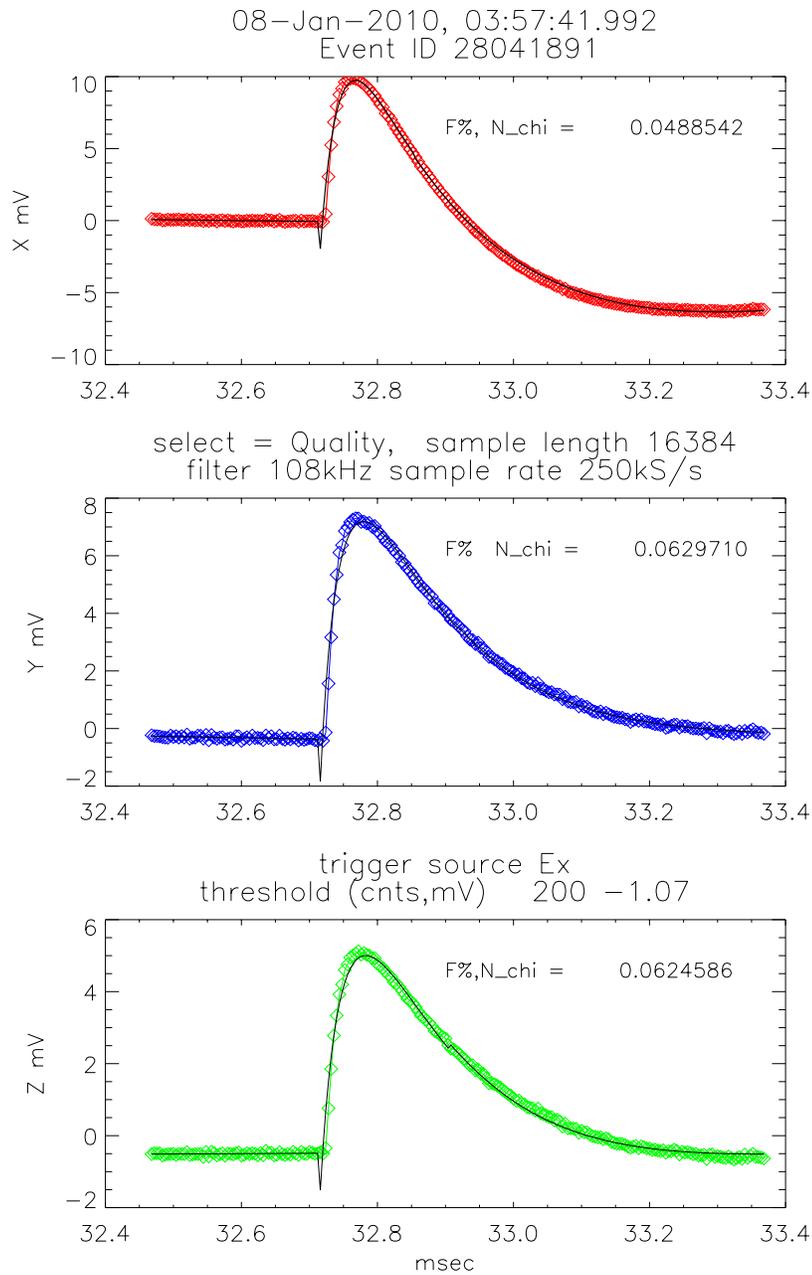

1. The waveform of a dust impact as measured on the three STEREO antennas.

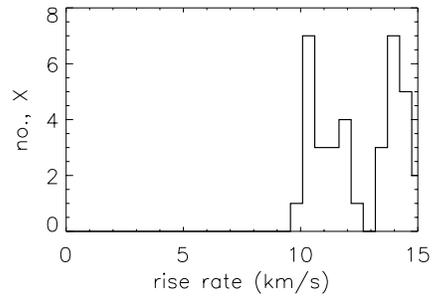
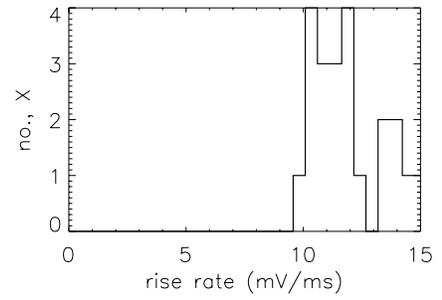
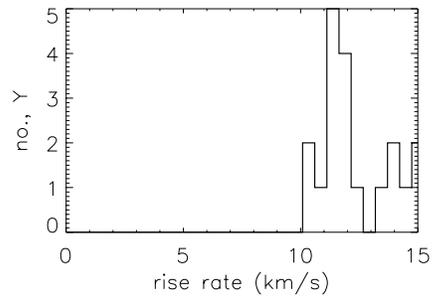
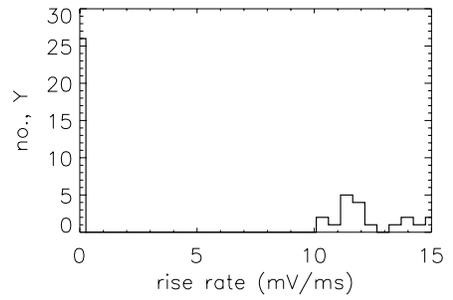
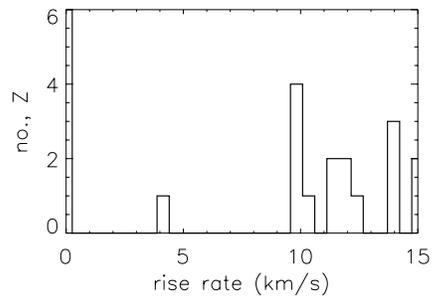
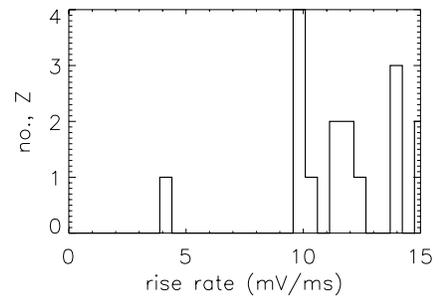

3. Histogram of rise speeds. See text for conversion from potential to rise in km/s.

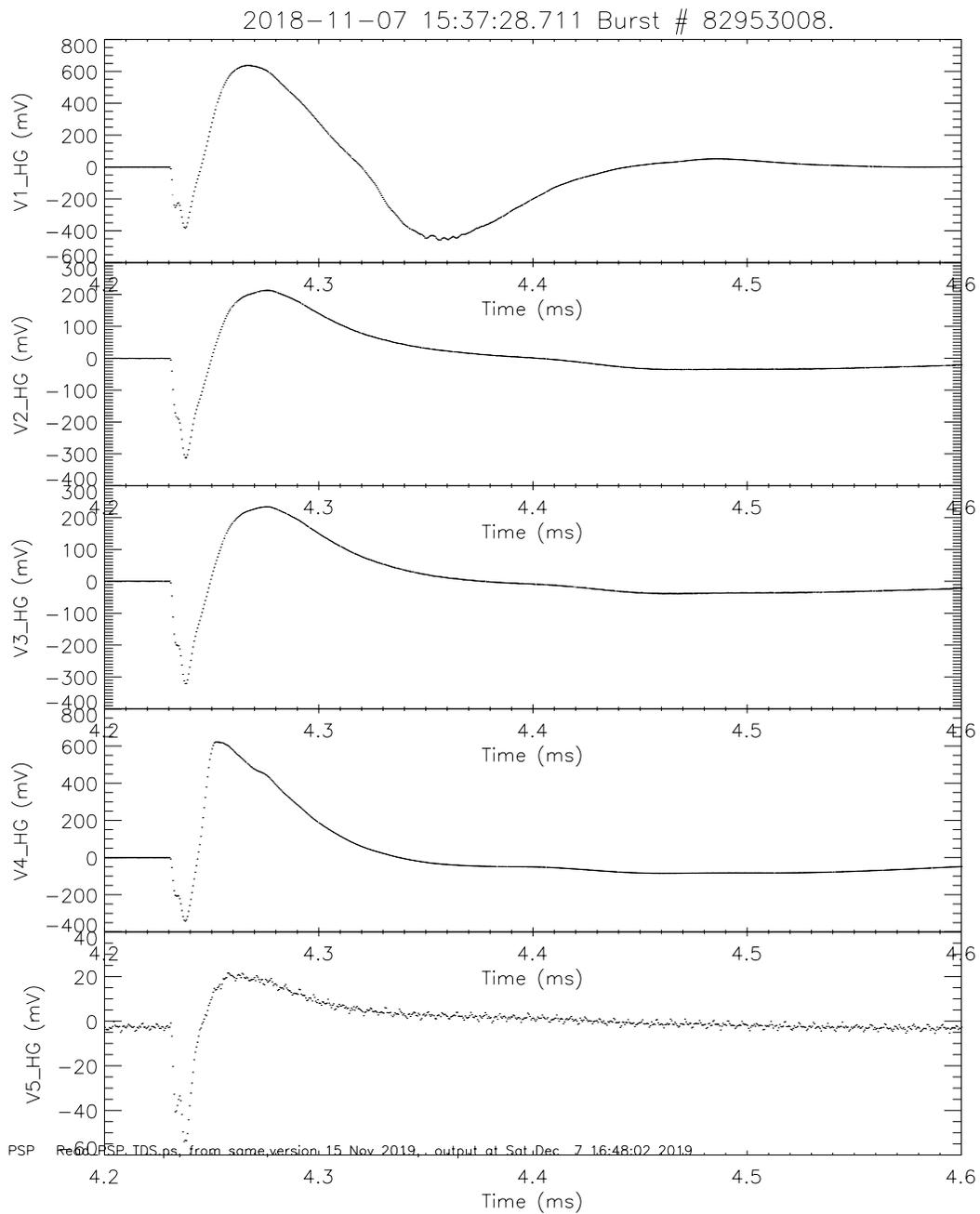

4. Waveform of a dust impact on Parker Solar Probe near first perihelion at 35 Solar radii. 6.

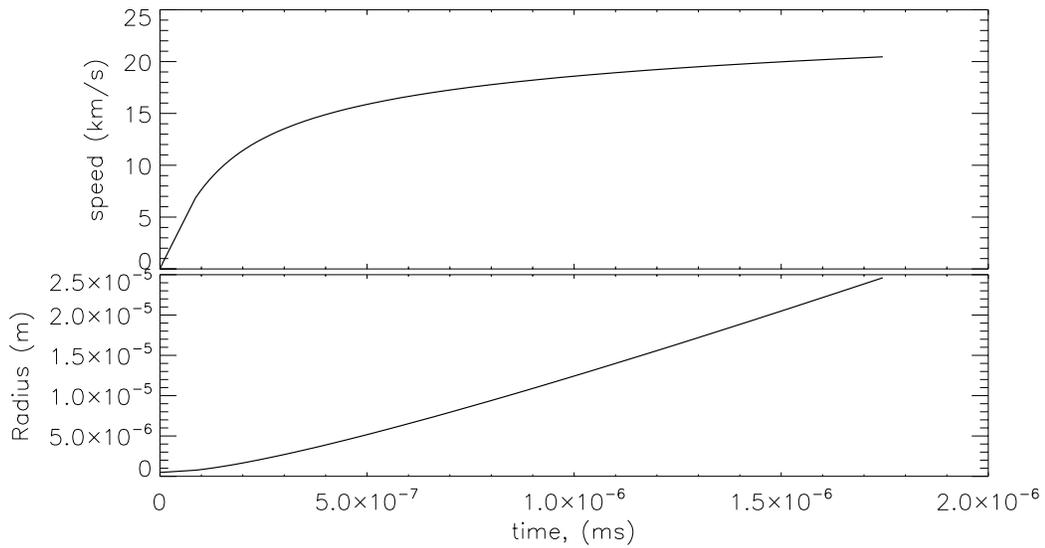

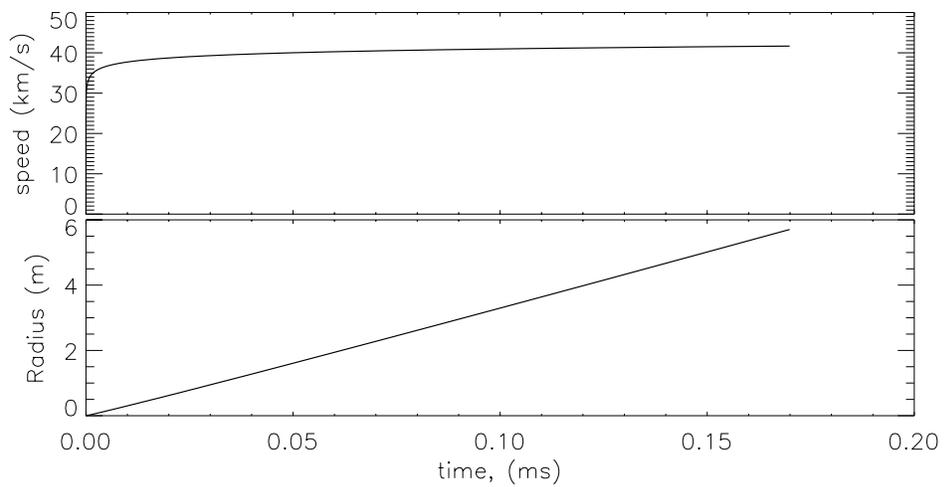

- 
- 5. Expansion speed due to electrostatic repulsion. The upper panel shows the early acceleration    and the lower panel shows expansion to about 5 meters radius.

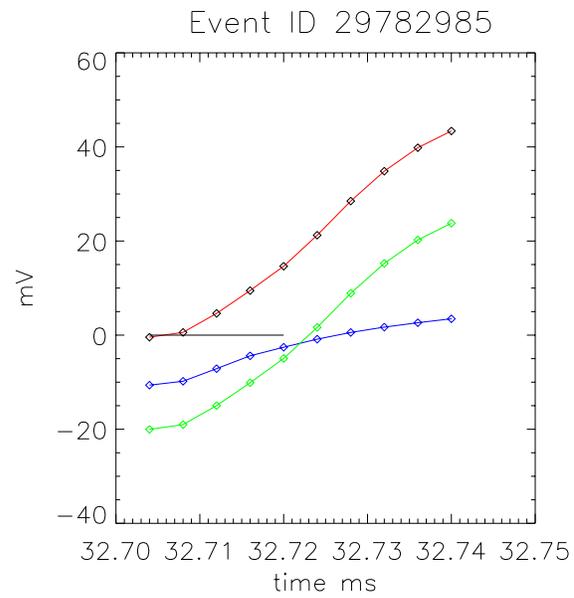

6. Acceleration due to the spacecraft electric field acting on the positive charge of the cloud, an event on Ahead at 2010/01/17 2113.

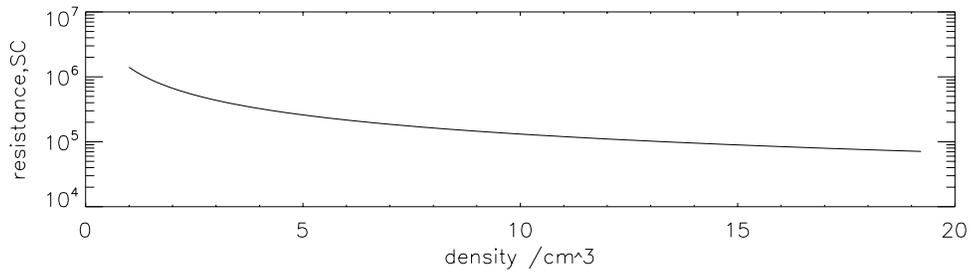

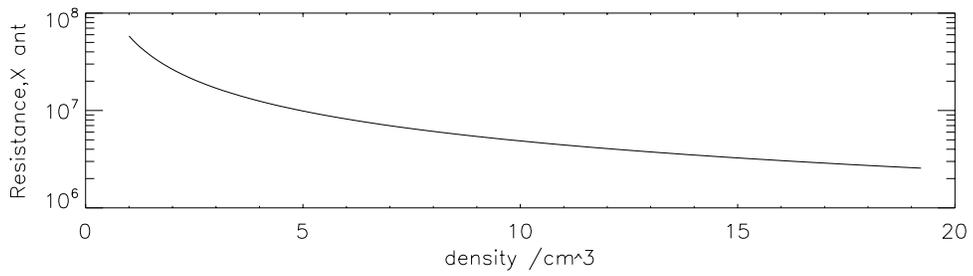

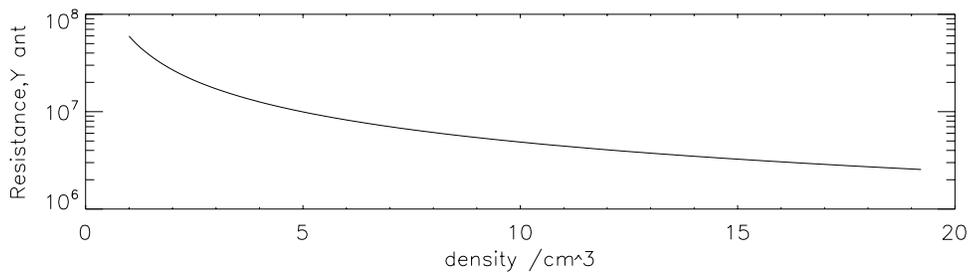

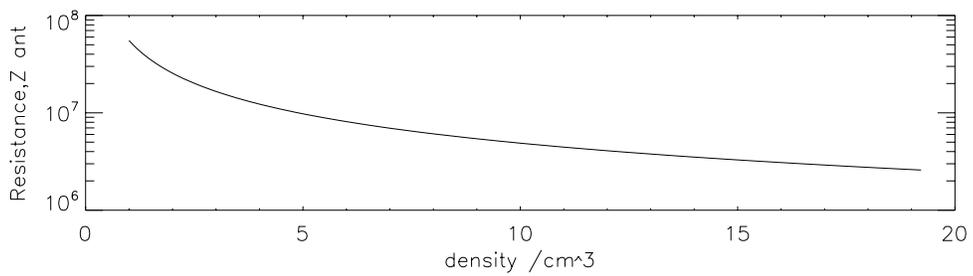

7. Calculations of the spacecraft-plasma and antenna-plasma resistance as a function of ambient plasma density.

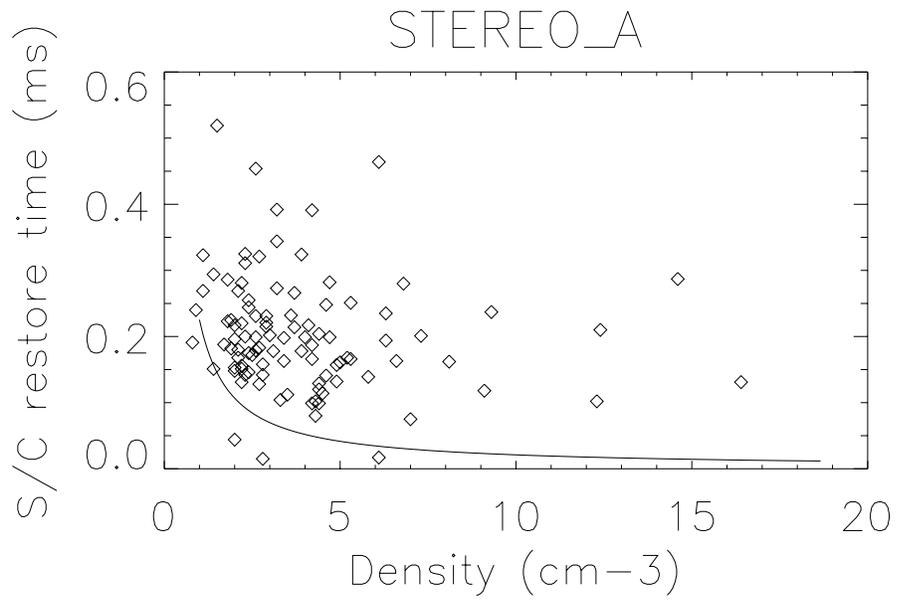

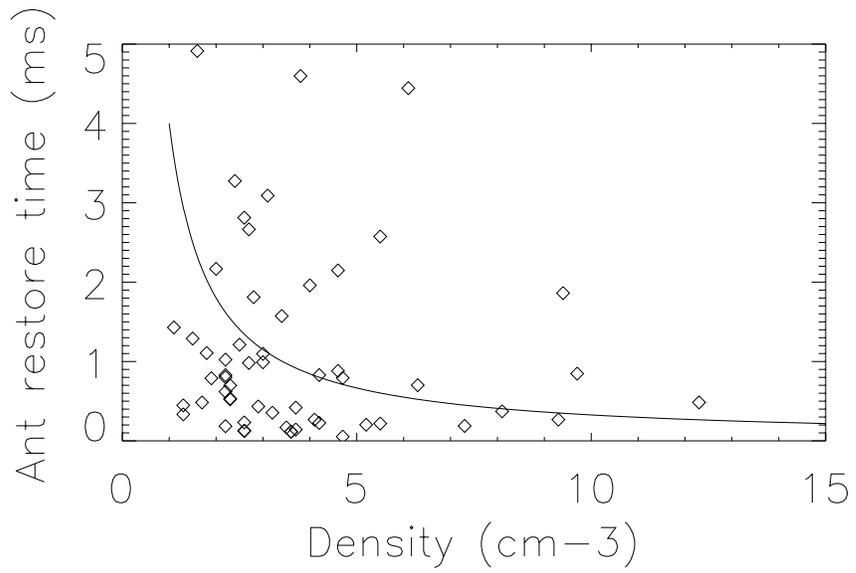

8 Restore times as functions of density from the resistance calculations shown in Figure 7.

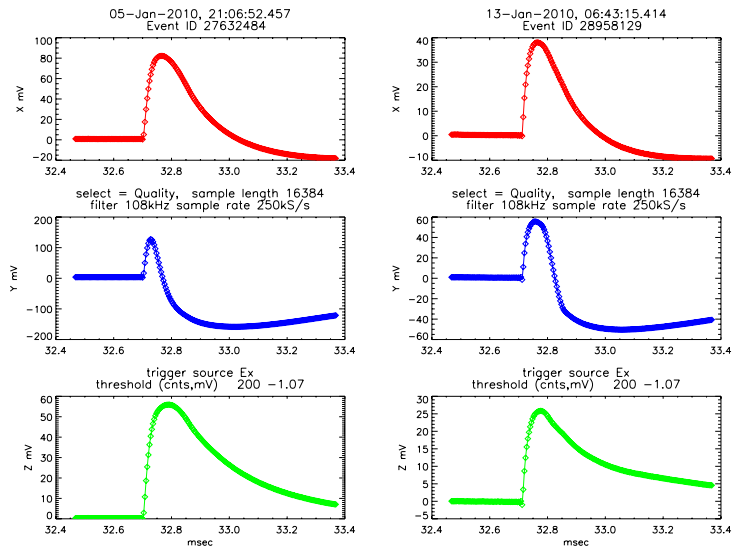

9. Two STEREO events thought to show deposition of electrons on an antenna

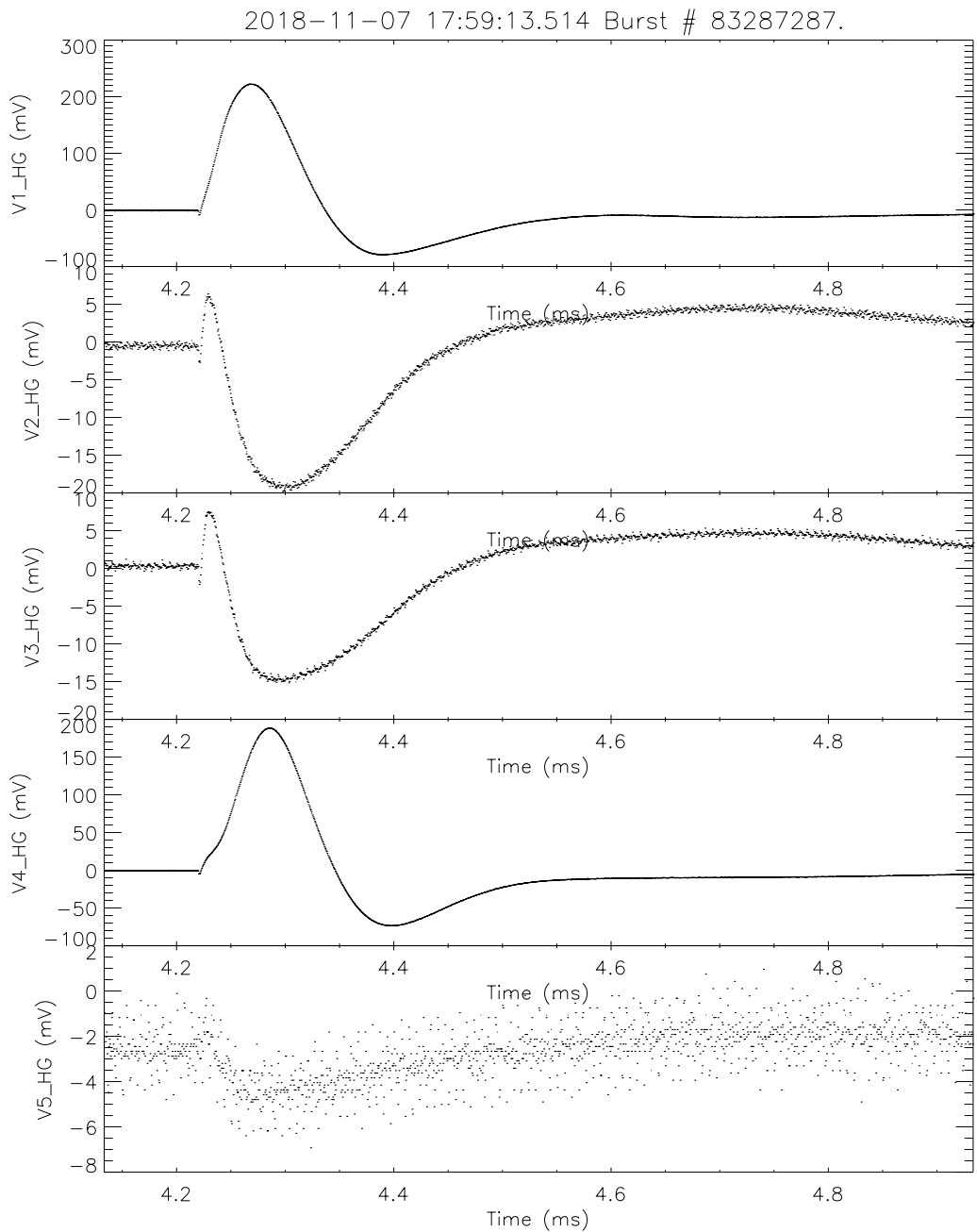

10. A similar event from the first perihelion of Parker Solar Prob.